\documentclass[final,5p,times,twocolumn,authoryear]{elsarticle}

\usepackage{lineno,hyperref}
\usepackage{mhchem}
\usepackage{natbib}
\usepackage{graphicx} 
\usepackage{booktabs}
\usepackage{multirow}
\usepackage{hhline}
\usepackage{xspace}
\usepackage{float}
\usepackage{mhchem}
\usepackage{subcaption}
\usepackage{lscape}
\usepackage{rotating}
\journal{Journal of \LaTeX\ Templates}
\usepackage[dvipsnames]{xcolor}
\colorlet{LightForestGreen}{ForestGreen!30!}
\colorlet{LightRubineRed}{RubineRed!30!}
\colorlet{LightBlue}{Blue!20!}
\newcommand{\HF}{HF\xspace}

\usepackage{soul}
\usepackage{latexsym}

\biboptions{authoryear}

\begin{document}

\begin{frontmatter}

\title{High level \textit{ab initio} binding energy distribution of molecules on interstellar ices: Hydrogen fluoride}
\tnotetext[mytitlenote]{Corresponding author, stvogtgeisse@qcmmlab.com}

\author[1]{Giulia Bovolenta}
\author[2]{Stefano Bovino} 
\author[1]{Esteban V\"ohringer-Martinez} 
\author[1]{David A. Saez} 
\author[3,4]{Tommaso Grassi} 
\author[1]{Stefan Vogt-Geisse\tnoteref{mytitlenote}$^{\ast,}$}

\address[1]{Departamento de F\'isico-Qu\'imica, Facultad de Ciencias Químicas, Universidad de Concepci\'on, Concepci\'on, Chile}
\address[2]{Departamento de Astronom\'ia, Facultad Ciencias F\'isicas y Matem\'aticas, Universidad de Concepci\'on,
Av. Esteban Iturra s/n Barrio Universitario, Casilla 160, Concepci\'on, Chile}
\address[3]{Universit\"ats-Sternwarte M\"unchen, Scheinerstr.~1, D-81679 M\"unchen, Germany} 
\address[4]{Excellence Cluster Origin and Structure of the Universe, Boltzmannstr.~2, D-85748 Garching bei M\"unchen, Germany}

  \begin{@twocolumnfalse}
\begin{abstract}
The knowledge of the binding energy of molecules on astrophysically relevant ices can 
help to obtain an estimate of the desorption rate, i.e. the molecules residence time on the 
surface. This represents an important parameter for astrochemical models, and it is crucial 
to determine the chemical fate of interstellar complex organic molecules formed on the surface 
of dust grains and observed in the densest regions of the interstellar medium through rich
rotational lines. In this work, we propose a new robust procedure to study the interaction 
of atoms and molecules with interstellar ices, based on \textit{ab initio} molecular dynamics 
and density functional theory, validated by high-level \textit{ab initio} methods at 
a CCSD(T)/CBS level. We have applied this procedure to a simple but astronomically relevant
molecule, hydrogen fluoride (HF), a promising tracer of the molecular
content of galaxies. In total we found 13 unique equilibrium structures of HF binding to small
water clusters of up to 4 molecules, with binding energies ranging from 1208 to 7162 K (2.4 to 14.23 kcal $\mathrm{mol^{-1}}$).
We computed a 22-molecules model of amorphous solid water (ASW) surface using \textit{ab initio}
molecular dynamics simulations and carried out a systematic analysis of the binding sites 
of HF, in terms of binding modes and binding energies. Considering 10 different 
water clusters configurations, we found a binding energy distribution with an average value of $5313\pm74$ K ($10.56\pm0.15$ kcal $\mathrm{mol^{-1}}$), and a dispersion of $921\pm115$ K ($1.83\pm0.23$ kcal $\mathrm{mol^{-1}}$). Finally, the effect of the electrostatic field of the 22 water 
molecules on the binding energies was investigated incrementally by symmetry adapted 
perturbation theory, in order to gauge the effect of the water environment on the binding
energies.  The results indicate that the extent of the electrostatic 
interaction of HF with ASW  depends strongly on the properties of the binding site on the water cluster. We 
expect that this work will provide a solid foundation for a systematic development of a 
binding energy distribution database of small molecules on astrophysically relevant surfaces.

\end{abstract}

 \end{@twocolumnfalse} \vspace{0.6cm}


\end{frontmatter}


\section{Introduction}
\noindent Dust is a key element for the formation and destruction of interstellar molecules, 
due to its catalytic nature  \citep{charnley2001,caselli2012}. In the dense and cold regions 
of the interstellar medium (ISM), dust grains are covered with ice mantles, formed by the
accretion of gas-phase atomic and molecular material onto the surface of bare grains (mainly
carbonaceous and silicates). Observational data provide a number of constraints both on their chemical composition and physical structure,
although precise 
information is still lacking  \citep{bartels-rausch2012}. Water is the primary 
ice component, along with significant quantities of other molecules -- mainly CO, NH$_3$, 
and \ce{CO2}  \citep{garrod2011, boogert2015}. Species adsorbed on the surface of these icy mantles are 
supposed to diffuse and lead to the formation of new molecules via thermal hopping or 
tunneling between different sites. This has been so far suggested to be one of the main
paths leading to the formation of interstellar complex organic molecules (iCOMs)
\citep{charnley2006,sewilo2019,simons2020}. Thermal or cosmic-rays induced desorption of 
the new species allows to observe them in gas-phase via rich rotational lines. Most of the parameters 
causing species desorption from grains are directly or indirectly related to their binding
energy ($E_b$) and, in addition, experiments have shown that $E_b$ is 
heavily dependent on the type of grain surface and other morphological parameters 
(e.g. coverage, \citealp{he2016}). An accurate knowledge of the binding energy is therefore of
central importance for the understanding of key chemical processes leading to the formation of Solar-type systems and life-precursor species.\\ 
Binding energies can be experimentally measured using Temperature Programmed Desorption (TPD)
-- a list can be found in the review by \citet{cuppen2017} -- but current facilities are affected
by sensitivity problems, identification of volatiles with the same mass and, more important, by 
the incapability to study radicals species. It is then necessary to complement experimental 
studies with high-accuracy theoretical calculations. \\
A reasonable choice for the theoretical description of the substrate is vapor-deposited amorphous
solid water (ASW), since at the typical temperatures of the interstellar cold regions, water is assumed
to be amorphous \citep{bartels-rausch2012, hama2013} and partly porous \citep{palumbo2006}. 
It should be kept in mind, though, that the simulations of this kind of structures 
present challenges related to the computational method and the model employed to describe the
surface.  Recently, there have been efforts to consider more sophisticated ASW models which 
take into account a variety of sites available for the adsorbate to bind to the surface. 
Among them, \citet{Shimonishi2018} proposed a mixed approach, where a large
amorphous surface is generated through molecular dynamic annealing simulations, followed by 
quantum chemistry  binding energies evaluation, in order to provide a new sets of values 
for the system C, N, \ce{O -ASW}. In addition, several reactivity studies have been carried 
out by using a QM/MM ASW 
simulated surface, including binding sites and binding energies analysis \citep{song2016,lamberts2019, molpeceres2020}.  The efforts to 
obtain a binding energy extensive catalogue for small molecules on ice surfaces
have been so far limited to Density Functional Theory (DFT) calculations on small water 
clusters (up to 6 molecules, \citealp[see][]{sil2017,das2018}) or interaction with water monomer by 
linear semi-empirical models \citep{Wakelam2017}, which do not capture the complete
statistical nature of the interaction on ASW.  Furthermore, closed shell molecules interact 
with water molecules through non-covalent interactions, which on average are just of the order 
of a millionth of the total molecular energy. Therefore, it is paramount to carefully validate 
any DFT methods through more accurate wavefunction approaches, in order to obtain reliable 
binding energies.

In this paper we present a new pipeline for the accurate computation of atomic and molecular
binding energies on astrophysically relevant surfaces, by employing \textit{ab initio} 
molecular dynamics  and DFT, validated by high-level wavefunction methods. We apply this 
procedure to a simple but astrophysically relevant species, hydrogen fluoride (\HF), 
a reliable tracer of the molecular gas in galaxies  \citep{neufeld2005}.  Fluorine is one of 
the few atoms which can undergo exothermic reaction with \ce{H2} to form a diatomic hydride, 
a reaction that has been proven to be a uniquely efficient pathway to \HF formation. Astronomical
\HF constitutes the main reservoir of interstellar fluorine, carrying almost the totality 
of the elemental fluorine \citep{neufeld2005}. In fact, the paths leading to \HF
destruction are not particularly efficient and the next most abundant molecule containing 
fluorine, \ce{CF^+}, is expected to show an abundance of two orders of magnitude smaller 
than that of \HF \citep{liszt2015,vanderwiel2016,muller2016}. Thus, the constant 
\ce{HF/H2} ratio, together with a high probability that \HF molecules are in the rotational 
ground state, results in a direct connection between \HF absorption depth and \ce{H2} column
density. The observational data, however, suggest that in cold clouds, where
temperatures are low ($<$ 20 K), \HF may condense onto dust grains.  Although freeze-out 
effects might reduce the effectiveness of HF as a diagnostic tool, the density and temperature
conditions needed for its adsorption, and desorption, have been studied in astrophysical 
context only qualitatively \citep{neufeld2005,vanderwiel2016}. 
Here we report the first high level \textit{ab initio} binding energy calculations of the 
system HF -- water (W) cluster up to 4 water molecules. These values are used as reference 
for the subsequent calculations involving larger ASW clusters consisting  of 22 water molecules.
Our ASW surface model has been obtained starting from \textit{ab initio} molecular dynamics 
(AIMD) and our procedure is designed to be applicable to other systems.
This paper is organised as follow: The computational methods are presented in Section \ref{sec:section2}, 
the computed structures and energies are presented in Section \ref{sec:section3} and 
discussed in Section \ref{sec:section4}. We then summarise and present our conclusions 
in Section \ref{sec:section5}.

\section{Computational Methods}\label{sec:section2}

\begin{figure}[ht]
    \centering
    \includegraphics[angle=0, width = 0.8\linewidth]{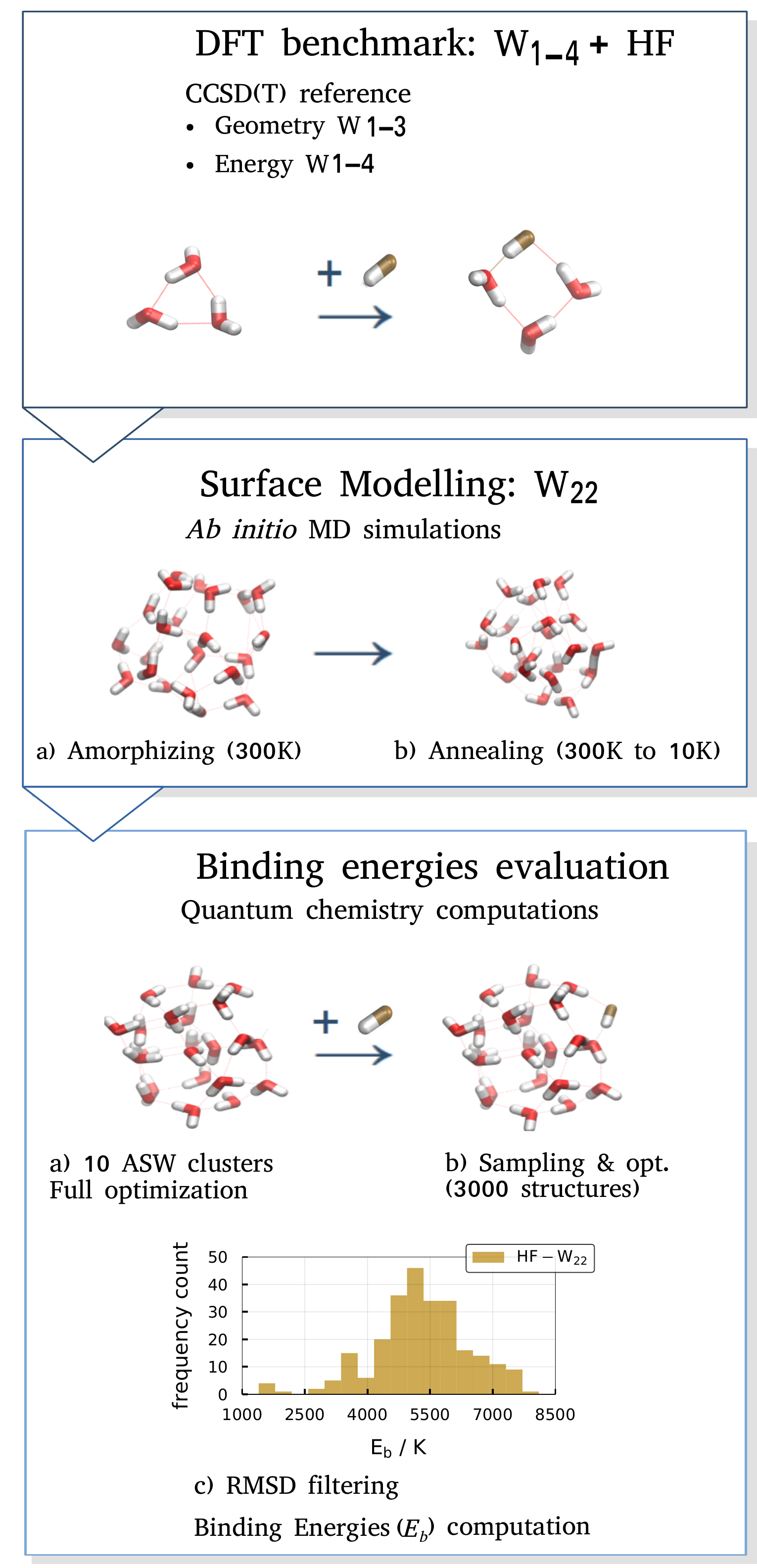}
    \caption{Three-step computational procedure used in this work for system simulation and binding energy evaluation. The color scheme for the
atoms is yellow for F, red for O, and white for H.}
    \label{fig:flow_chart}
\end{figure}

\noindent The computational pipeline used in this work is displayed in Fig. \ref{fig:flow_chart}.
After a first benchmark step (1), aimed at predicting suitable DFT methods for the different 
phases of the procedure, the amorphous water cluster is modelled (2) and finally the binding 
sites of HF on the surface are identified and the corresponding binding energies computed (3).
 
\subsection{DFT benchmark}
\noindent We carried out geometry and energy benchmarks considering 45 different DFT levels of
theory, dividing them in three tiers of computational cost and accuracy: 
1) Generalized Gradient Approximation (GGA)/def2-SVP; 2) Hybrid-GGA/def2-TZVP; 
and 3) Meta-GGA/def2-TZVP \citep{weigend2005a}. Since DFT alone fails to correctly describe
dispersion interactions, the D3BJ \citep{grimme2010,grimme2011} dispersion correction was 
included in the geometry optimizations and single point computations.   All benchmark 
computations were done using the \textsc{Psi4} program package \citep{parrish_psi4_2017}.
\subsubsection*{Geometries}
\noindent The water clusters equilibrium structures, \ce{W_{1-3}}, are obtained by geometry
optimizations using a wavefunction at the CCSD(T) level \citep{raghavachari1989}, together 
with the correlation consistent valence polarized aug-cc-pVTZ basis 
\citep{DunningJr1989,kendall1992}. To find the reference structures for the interacting
\ce{HF-W_{1-3}} systems, we employ a random minima search. The initial exploration is
done at the BLYP/def2-SVP level of theory and the resulting equilibrium
structures are optimized using first MP2/cc-pVTZ followed by DF-CCSD(T)/aug-cc-PVTZ method 
and basis \citep{bozkaya2017}.
HF -- \ce{W_4} equilibrium geometries, used in the energy benchmark, have been computed 
at DFT revPBE0/def2-TZVP level, due to computational cost, and are not part of the geometry reference structures.

{}

\subsubsection*{Energies}
\noindent The binding energies of the reference water cluster model systems, HF -- \ce{W_{1-4}}, 
are computed with the CCSD(T) method at the complete basis set (CBS) limit
\citep{klopper1986,feller1992,helgaker1997,karton2006}, by employing the following equation:

\begin{equation}\label{eq:BE_terms}
 E_b = E_{sup} - (E_{\mathrm{W}} + E_{\mathrm{ads}}) 
\end{equation}

\noindent where $E_{sup}$ stands for the energy of the HF -- \ce{W_n} supermolecule, $E_{W}$ is
referred to the \ce{W_n} cluster energy and $E_\mathrm{ads}$ is the energy of the adsorbate. 
The zero point vibrational energy  (ZPVE) corrections to the binding energies are obtained 
at the CCSD(T)/aug-cc-pVTZ level of theory.  Furthermore, to correct for basis set 
superposition error (BSSE) of the  def2-SVP and def2-TZVP basis sets used at the DFT level 
of theory,  the counterpoise correction \citep{boys1970} is added to both the $E_b$ and 
the $\Delta_{ZPVE}$ computations. Further details and benchmark tables can be found in
Supplementary Material (Fig. S1-S12).  

\subsection{Surface Modelling: \textit{Ab initio} molecular dynamics simulations}\label{sec:surfmodel}
\noindent The modelling procedure aims at generating an amorphous water surface. To build the ASW ice
model we  use the \textit{cluster approach} \citep{zamirri2019} by selecting a suitable
cluster size (22 water molecules), in order to guarantee a reasonable number of available 
binding sites on the surface, while at the same time being able to use high-level model 
chemistry. The steps to obtain the ASW ice model are the following:

1) First, we perform a high temperature \textit{ab initio} molecular dynamics (AIMD) simulation 
of 100 ps on the initial system in order to amorphyze it. 
We preferred AIMD over classical dynamics for ice simulations,
since the former generates interaction potentials using quantum-chemical methods, while common
empirical water force fields  are parameterized using properties of liquid water \citep{szalewicz2009, chen2017}.  
(AIMD parameters: 1 fs for the time-step, Langevin
thermostat at 300 K, spherical periodic boundary conditions applied in order to avoid evaporation
effects);

2) We extract 100 independent structures  ($\tau_{correlation} \simeq$ 1 ps) from the resulting trajectory, which undergo 
temperature annealing of 3 ps to reach the target interstellar conditions (10 K); 

3) Among the structures obtained, the first 10 more representative are selected using 
geometrical criteria (we assumed a similarity threshold of root-mean-square deviation of atomic
positions (RMSD) $\leq$ 0.4 \text{\AA} to group the obtained structures and selected the 10 most populated clusters). 
All AIMD simulations are performed at the BLYP/def2-SVP level, adding 
D3 Grimme  \citep{grimme2010} correction for dispersion interactions, as implemented in
\textsc{Terachem}  \citep{ufimtsev2009, titov2013}.

\subsection{Binding sites and binding energies evaluation}\label{sec:bindingsites}
\noindent The previously selected 10 ASW clusters, fully optimized at BLYP/def2-SVP level, 
present an amorphous spheroidal structure with an approximate size of 
$(6 \times 5 \times 7$) \AA$^3$. The target molecule is added to each cluster using a random
surface sampling procedure. 3000 structures (300 per cluster) are generated and optimized at
BLYP/def2-SVP level of theory.  Subsequently, the unique resulting structures are filtered using
geometrical criteria (RMSD = 0.6 \text{\AA}) , and serve as  initial candidates for molecule HF -- ASW binding sites.\\ 
We refine the geometry at the revPBE0/def2-TZVP \citep{zhang1998} level of theory, 
leading to the generation of 255 equilibrium structures in which the molecule is bound to 
the ASW surface through different sites. Finally, the binding energies are obtained using the 
same level of theory. Given the high variety of binding sites on the amorphous surface, 
we collect a set of energies for each of the 10 clusters. Assuming that the clusters share 
the same morphological characteristics, since they have been generated from a unique AIMD
trajectory and annealed in the same way, the binding energies collected are considered to
constitute a single binding energy distribution of the target molecule on ASW.\\ 
The \textsc{Terachem} program was used for the initial minima search DFT optimizations at the 
lower GGA level of theory and \textsc{Psi4} was used for the hybrid GGA computations.

\subsection{Binding energy analysis}

\begin{figure*}[ht!]
    \centering
    \includegraphics[angle=0, width = 0.8\linewidth]{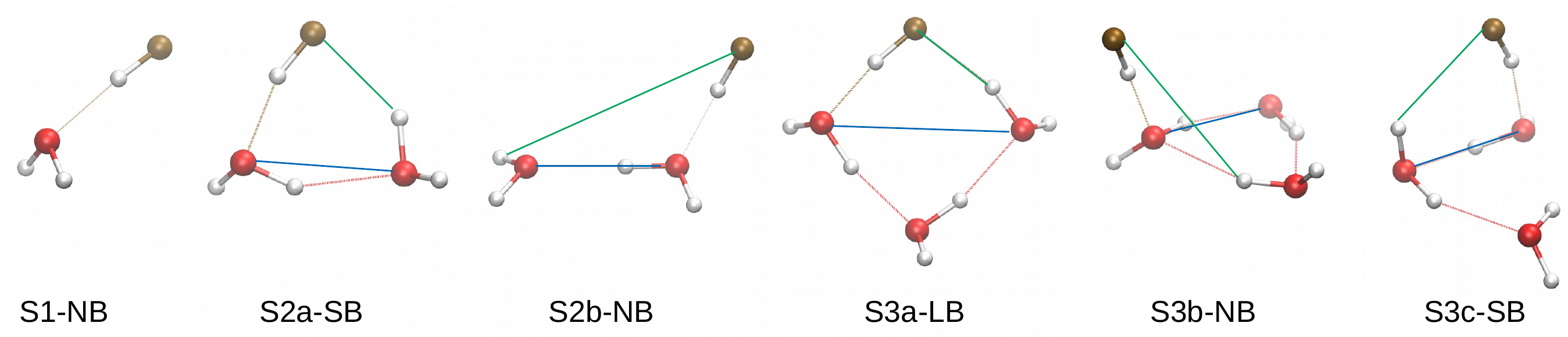}
    \caption{CCSD(T)/aug-cc-pVTZ equilibrium geometries found for HF -- W systems up to 3 water
    molecules. For each structure the distance between the fluorine atom and the closest water
    cluster hydrogen belonging to a different water molecule ($R_{F-Hw}$) is highlighted in green;
    the distance between the two oxygen atoms closest to the fluorine atom ($R_{O-O}$) is
    highlighted in blue. The bond lengths are shown in Table \ref{tbl:small clusters}.  
    The color scheme for the atoms is yellow for F, red for O, and white for H.}
    \label{fig:small clusters}
\end{figure*}{}

\noindent The binding energy can further be decomposed into two components: deformation energy
($E_{def}$) and interaction energy ($E_{int}$),  such that:

\begin{equation}\label{eq:IE_terms}
 E_b = E_{int} - E_{def} 
 \end{equation}
The interaction energy accounts for the non-covalent interaction of the two molecular
fragments in the bound conformation. The deformation energy quantitatively characterises 
the structural changes of the adsorbate and the water cluster with respect to the isolated
molecules and is defined by:

\begin{equation}\label{eq:IE_terms}
 E_{def} = (E_{ads}^{sup} + E_{{W}}^{sup}) - (E_{ads}^0 + E_{{W}}^0),
 \end{equation}
 
\noindent where $E_{ads}^0$ and $E_{W}^0$ is the energy of the isolated molecules and $E_{ads}^{sup}$ 
and  $E_{W}^{sup}$ the energy of the adsorbate and water cluster in the supermolecular
configuration.

\noindent Finally, we perform a zeroth order Symmetry Adapted Perturbation Theory (SAPT0)  analysis
\citep{jeziorski1994}, by using a jun-cc-pVDZ  \citep{papajak2011} basis set on 
a few selected structures. This SAPT0 energy decomposition allows to further
decompose the interaction energy into electrostatic, induction, dispersion and exchange-repulsion, 
which is a useful tool for identifying the effects that drive the non-covalent interaction between 
the adsorbate and the ASW surface. Within this decomposition analysis, the total interaction
energy is given by:

\begin{equation}\label{eq:IE_terms}
 E_{int} = E_{elest} + E_{exch} + E_{ind} + E_{disp} 
 \end{equation}
 
\noindent All SAPT0 computations use the density-fitted implementation 
\citep{hohenstein2010, hohenstein2011} provided in \textsc{Psi4}.

\section{Results}\label{sec:section3}
\noindent In the first part of this section, we report the results of the highly accurate 
HF -- \ce{W_{1-3}} geometries and HF -- \ce{W_{1-4}} binding energies 
used as a reference in the DFT benchmark (see Sec. \ref{sec:dft_bench}).
Additionally, we computed DFT binding energies for the systems \ce{HF - W_{5-6}} 
(Sec. \ref{sec:five_W}).  Finally (Sec. \ref{sec:dft_ASW}), we report the structures and 
the binding energies of HF on ASW model surface, \ce{HF - W_22}.

\subsection{DFT benchmark}\label{sec:dft_bench}

\subsubsection{CCSD(T) equilibrium geometries}

\noindent Fig. \ref{fig:small clusters} shows the CCSD(T) \ce{HF - W_{1-3}} equilibrium geometries
employed as references for the DFT benchmark.  We found one isomer for the system of HF 
interacting with the water monomer, two for the water dimer, and three for the trimer.
These structures  differ on how HF is bound to the clusters: through two hydrogen bonds, where 
HF binds to two different water molecules forming a bridge-type interaction (e.g. S2a-SB,
\textit{short bridge} abv. SB, or S3a-LB, \textit{long bridge} abv. LB, depending on the
oxygen-oxygen length), or through a single hydrogen bond with only one water molecule 
(e.g. S2b-NB, \textit{non-bridged} abv. NB). This classification, albeit evident in the  
small water clusters, is very helpful when analyzing the \ce{HF - ASW} system, since it allows
for automatic identification and a separate energetic analysis of the different binding modes.
We explain the classification procedure in detail in section \ref{sec:section41}.
In terms of geometry, the global best DFT functional is the Hybrid-revPBE0 coupled with
def2-TZVP basis set, with an average RMSD of only 0.020. 
The complete set of benchmark
tables is reported in the Supplementary Material (Fig. S1-S12).
Since the performance of BLYP is similar (average RMSD 0.175) to HCTH120 (average RMSD 0.157), 
we preferred to rely on it for the computations carried out at GGA level, due to 
its more widespread implementation in electronic structure programs.
It is worth mentioning that the CCSD(T)/aug-cc-pVTZ minima search of the S3b-NB structure was an
onerous endeavour, which indicates a very flat potential energy surface around this equilibrium
geometry. Furthermore, not surprisingly, several DFT functionals failed to correctly optimize 
this structure.

\subsubsection{CCSD(T) binding energies}

\begin{table*}[h!]
     \centering
     \small
     \caption{The binding energy ($E_b$) including BSSE counterpoise correction for \ce{HF - W_{1-4}} systems. The zero point vibrational energy ($\Delta_{ZPVE}$) correction is added when indicated and it is counterpoise corrected. The table also reports deformation energy ($E_{def}$) and interaction energy ($E_{int}$). All the energy values are in Kelvin. The last two columns show geometrical parameters relative to the binding sites, see Fig.\ref{fig:small clusters}. The structures' names are referred to the classification made in section \ref{sec:section41}. List of abbreviation used: SB -- \textit{short bridge}, LB -- \textit{long bridge}, NB -- \textit{not-bridged}, Y -- Y type, Avg -- average. Energy values in kcal $\mathrm{mol^{-1}}$ are reported in Supplementary Material}
     \label{tbl:small clusters}
      \begin{tabular*}{\textwidth}{@{\extracolsep{\fill}}l l  c c c l l l l l }
      
\hline
\multicolumn{2}{c}{\multirow{1}{*}{\ce{HF - W_n}}} &  \multicolumn{2}{c}{CCSD(T)/CBS} & \multicolumn{6}{c}{revPBE0/def2-TZVP}  \\\cmidrule{3-4} \cmidrule{5-10}
 n & Name  & \multicolumn{1}{c}{$E_b$} & \multicolumn{1}{c}{+$\Delta_{ZPVE}$} & \multicolumn{1}{c}{$E_b$} & \multicolumn{1}{c}{+$\Delta_{ZPVE}$} & $E_{def}$  & $E_{int}$ & \multirow{1}{*}{$R_{F-Hw}$ / \text{\AA}} & \multirow{1}{*}{$R_{O-O}$ / \text{\AA}} \\
 \hline
 \multirow{1}{*}{1} &  S1-NB & 4479 &   3090 & 4390   & 3074   & 9 & 4399 & -  & - \\
 \hline
  \multirow{4}{*}{2}  & Avg &  4684 & - & 4622 &  3420     & 294 & 3714 & 2.39 & 2.83 \\
                   & S2a-SB   & 6799 & 5003 & 6667 &   4925 &  343 & 7010 & 2.08  & 2.76  \\
                   & S2b-NB  &  5355 &  4131 & 5286 & 4118 &  507 & 5793 & 5.16  & 2.85  \\
                   & S2c-NB  & 1898 &   - & 1913   & 1372   & 31 & 1944 & 4.06  & 2.89   \\
\hline
 \multirow{4}{*}{3} & Avg & 5414 &  - & 5234  &  4185    & 1423 & 6657 & 2.54 & 3.12 \\
 
                    & S3a-LB  & 7162  &  5764 &  7172   & 5907    & 2402 & 9574 & 1.82      & 3.76  \\
                   
                    & S3b-NB     & 4786   & - & 4031  & 3184                   & 938 & 4969 &  4.50     & 2.91   \\ 
                    & S3c-SB    & 4580 & - & 4500   & 3440                  & 928 & 5427 &  2.61     & 2.68      \\
\hline
 
 \multirow{7}{*}{4} & Avg &   4957              & - &   4886 &  3805  & 1861 & 6747 & 2.30  & 2.98  \\
                  & S4a-LB      & 6211   & - & 6103        &  4974       & 4507 & 10610 & 1.76  & 4.26  \\ 
                  & S4b-LB      & 6208      & - &  5989    &   4575      & 2927 & 8915 & 1.86    & 3.52  \\
                  & S4c-SBY  &  5646  &    - & 5671         &   4416  &   1799 & 7471 &  2.38 &   3.01  \\
                  & S4d-SB       & 5338  & -    &  5229     &   4052     & 1841 & 7070 &    2.10       & 3.03  \\
                  & S4e-NB       & 5132        &  - &  5116 &   4004    & 19 & 5134 & 3.70     & 2.64  \\
                  & S4f-NB   & 1208 &   - &  1211            &    806   &       74 & 1285 &  4.08  &   2.72  \\

\end{tabular*}
\end{table*}

\begin{table*}[h!]
\small
     \centering
     \caption{The average binding energy ($E_b$) including BSSE counterpoise correction, with and without zero point vibrational energy ($\Delta_{ZPVE}$) correction, along with deformation energy ($E_{def}$) and interaction energy ($E_{int}$), for \ce{HF - W_{5-6}} systems. All the energy values are in Kelvin. The Type column is referred to the classification made in section \ref{sec:section41}. The last two columns show geometrical parameters relative to the binding sites, see Fig.\ref{fig:small clusters}. List of abbreviation used: SB -- \textit{short bridge}, LB -- \textit{long bridge}, NB -- \textit{not-bridged}, Y -- Y type. The error of the reported binding energies are in the range of the MAD obtained from the energy benchmark (74K).
     Energy values in kcal $\mathrm{mol^{-1}}$ are reported in Supplementary Material}
     \label{tbl:W5_W6}
      \begin{tabular*}{\textwidth}{@{\extracolsep{\fill}}l l l  l l l l l l}
\hline
 \multicolumn{2}{c}{{HF -- \ce{W_n}}} & & \multicolumn{6}{c}{revPBE0/def2-TZVP} \\ \cmidrule{4-9}   
  n & Type & num  & \multicolumn{1}{c}{$E_b$} & \multicolumn{1}{c}{+$\Delta_{ZPVE}$} & $E_{def}$ &  $E_{int}$ & \multirow{1}{*}{$R_{F-Hw}$ / \text{\AA}} & \multirow{1}{*}{$R_{O-O}$ / \text{\AA}} \\
\hline
 \multirow{6}{*}{5}  & Tot & 13  & 5579 &  4333 & 1739 & 7317 & 2.40 & 3.12 \\
  &  LBY &  1 &   6899          & 5421 &   2319 & 9217  & 2.14          & 3.87 \\  
  &  LB     &  4   & 6107    & 4836  & 2848 &  8955      & 1.95   & 3.61  \\ 
  & SBY   &  1 &    5962   & 4764 & 2756  &  8718    & 2.13        & 3.08    \\ 
  &  SB      &  6   & 4983  & 3774   & 968 & 5952   & 2.54   & 2.82  \\ 
  &  NB        &  1 &  5333    & 4154 & 326   & 5660      &  3.84      & 2.77 \\ 
 \hline
 \multirow{6}{*}{6}  & Tot & 30   & 6073 &  4609 & 1749 &  7822 & 2.21 & 3.49 \\
   &  LBY &   2  & 7753  &  5833 &  1477 & 9231 & 2.19 &  3.91  \\
   &  LB     &   15 &  6420 &  4956  & 2197 & 8617  & 1.97 &  3.99  \\ 
   & SBY   &   1 &   6121 &  4300  & 751  & 6873 &  2.41   &  2.80  \\ 
   &  SB      &   9 &  5525  &  4056 & 1581 & 7106 & 2.29 &   2.93  \\ 
   &  NB      &   3 &  4851  &  3819 & 525 & 5377   & 4.59 &  2.68  \\ 
 
\end{tabular*}
\end{table*}
\noindent In Table \ref{tbl:small clusters} we present the binding energy values obtained for
the HF -- \ce{W_{1-4}} systems. For the energy benchmark, we considered the CCSD(T) 
structures previously mentioned (Fig. \ref{fig:small clusters}), together with the 6 energy
minima of the HF -- \ce{W_4} system, optimized at the revPBE0/def2-TZVP level of theory.
Furthermore, we included a dimer equilibrium structure (S2c-NB) that we found only at DFT level.\\
The equilibrium structure with the highest binding energy is S3a-LB, with a $E_b$ value of 7162 K, followed by
S2a-SB (6799 K) and two HF -- tetramer minima (S4a-LB 6211 K and S4b-LB 6208 K), bridge-type as well. All high binding energy
equilibrium structures correspond to bridged type binding modes. The entire binding
energy range spreads from 1208 to 7162 K. The ZPVE correction have been computed at
CCSD(T)/aug-cc-pVTZ level only for four structures, due to computational cost, and it 
reduces the binding energies by $\sim$ 1300-1800 K for all of them.

The energy benchmark result is the following. 
Two Hybrid-GGA methods showed the best
performance: B97-2 (mean absolute deviation, MAD, 61 K) and revPBE0 (MAD 74 K). The complete set of benchmark tables is in Fig. S1-S12. It is worth mentioning that, 
due to the very flat potential energy surface (PES) around the S3b-NB minimum, this structure is the one
with the highest deviation from the CCSD(T) reference, especially when computing the 
binding energy on the DFT optimized structures (see Table\ref{tbl:small clusters}), which 
otherwise yields results in excellent agreement with the high level reference.
The DFT model chemistry that shows the highest fidelity with respect
to the CCSD(T) reference in terms of energy and geometry is revPBE0/def2-TZVP.
Therefore, we used this level of theory for the evaluation of binding sites and energies
in the   HF -- \ce{W_{5-6}} and the HF -- ASW$_{22}$ systems, in which high level coupled cluster
methods are prohibitively expensive. The ZPVE correction, computed 
at DFT revPBE0/def2-TZVP level, reduces the binding energies by around 1300-1600 K for all the
structures and maintains the good quality of the benchmark results.\\ \\ 

\begin{table*}[h!]
\small
     \centering
     \caption{The binding energy ($E_b$) including BSSE counterpoise correction for the system \ce{HF - ASW_22}. For the total (Tot) and for each category defined in section \ref{sec:section41}, we report the highest and the average binding energy values, with and without zero point vibrational energy ($\Delta_{ZPVE}$) correction. For the average values, we include deformation energy ($E_{def}$), interaction energy ($E_{int}$), and the standard deviation ($\sigma$). All the energy values are in Kelvin. The error of the reported binding energies are in the range of the MAD obtained
     from the energy benchmark (74K). List of abbreviation used: SB -- \textit{short bridge}, LB -- \textit{long bridge}, NB -- \textit{not-bridged}, Y -- Y type. Energy values in kcal $\mathrm{mol^{-1}}$ are reported in Supplementary Material
     }
      \label{tbl:BE_ASW}
      \begin{tabular*}{\textwidth}{@{\extracolsep{\fill}}l  l l l l l l l l l  l }
\hline
 \multicolumn{2}{c}{\multirow{2}{*}{HF -- \ce{ASW_22}}} &  \multicolumn{9}{c}{revPBE0/def2-TZVP} \\ \cmidrule{3-11} 
 
 & & \multicolumn{2}{c}{Highest values} & \multicolumn{5}{c}{Average values} & & \\ \cmidrule{3-4}\cmidrule{5-11}
  Type &  $\%$ & \multicolumn{1}{c}{$E_b$}  & \multicolumn{1}{c}{+$\Delta_{ZPVE}$} & \multicolumn{1}{c}{$E_b$}  & \multicolumn{1}{c}{+$\Delta_{ZPVE}$} & $E_d$ & $E_i$ & $\sigma$ &  \multicolumn{1}{c}{\multirow{1}{*}{$R_{F-Hw}$ / \text{\AA}}} & \multicolumn{1}{c}{\multirow{1}{*}{$R_{O-O}$ / \text{\AA}}} \\ 
\hline
    Tot   & 100.00   & 7841 & 6479 & 5263 & 3461  & 2012 & 7502   & 1181    & 2.15 & 3.52\\ 
    LBY   & 2.40     & 7418 & 6304 & 6534 & -     & 3286 & 9820  & 902    & 1.96 & 4.0  \\
    LB    & 47.06    & 7841 & 6479 & 5888 & 4898  & 2340 & 8228   & 967    & 1.97 &   3.94 \\
    SB    & 43.10    & 6425 & 5008 & 4659 & 3709  & 1151 & 5810   & 818     & 2.72 & 2.83 \\
    NB    & 7.44     & 5611 & 4491 & 4340 & 3957  &  504 &  4844   & 1575   & 4.68 & 2.75 \\
 \end{tabular*}
\end{table*}

\hfill \break
\subsection{HF -- \ce{W_{5-6}}}\label{sec:five_W}
\begin{figure}[ht]
    \centering
    \includegraphics[angle=0, width = 0.9\linewidth]{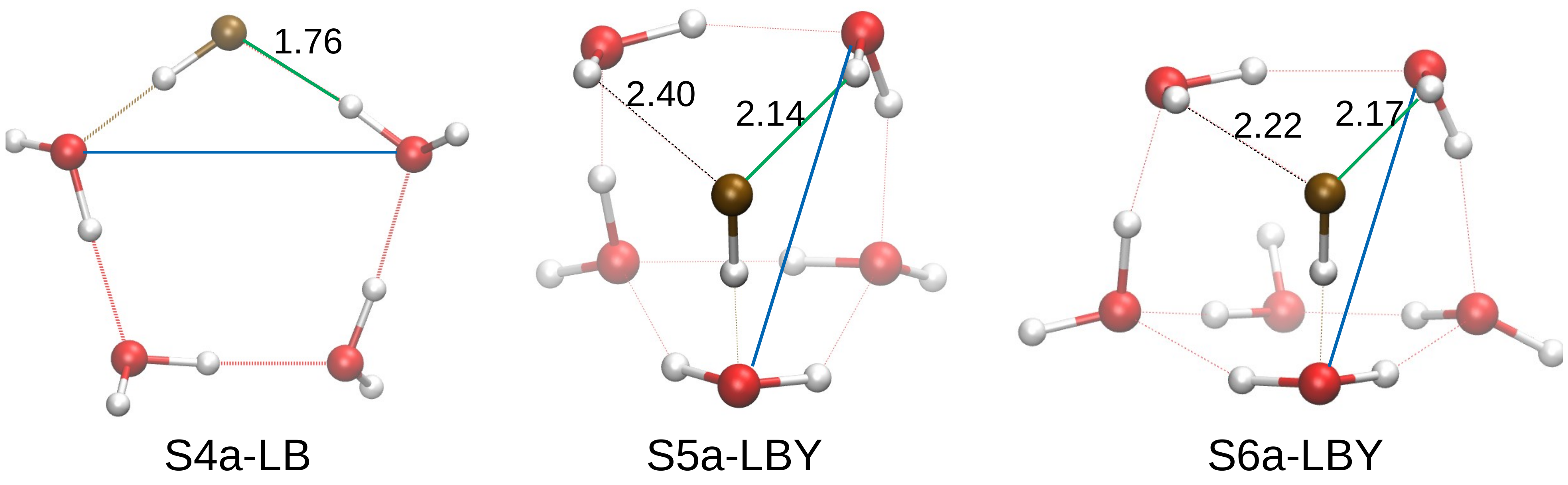}
    \caption{DFT revPBE0/def2-TZVP geometries corresponding to the global minima of \ce{HF - W_{4}} ($E_b$: 6103 K), \ce{HF - W_{5}} ($E_b$: 6899 K) and \ce{HF - W_{6}} ($E_b$: 8825 K)  systems. $R_{F-Hw}$ distances are underlined in green and $R_{O-O}$ in blue, see Fig.\ref{fig:small clusters}. The color scheme for the
atoms is yellow for F, red for O, and white for H. Interatomic distances
are given in \text{\AA}.}
    \label{fig:w5_w6}
\end{figure}

\noindent We studied the HF binding sites for the intermediate sized W$_{5-6}$  water 
clusters in order to obtain a more complete census of possible binding modes.
We found 13 equilibrium structures for \ce{HF-W_5} and 30 for \ce{HF-W_6}. 
The computed average $E_b$ for each binding mode (Table \ref{tbl:W5_W6}) 
ranges from 4851 to 7753 K (reduced by 1000-1500 K with ZPVE correction). 
Interestingly, in the global minimum structures depicted in Fig. \ref{fig:w5_w6}, fluorine 
forms a double hydrogen bond with two water molecules, with a $R_{F-Hw}$ distance of around
2.2 \text{\AA}, resulting in a "Y-like" shape. 
As expected, the spread in binding energy is higher in the hexamer (4500-7000 K) than in 
the pentamer (5300-6900 K).  This can be attributed to a richer variety of binding 
sites in the hexamer, which results in the emergence of a distribution (Fig. S16)
centered around 6000 K, with standard deviation of 1090 K. \\

{}

\subsection{HF -- ASW surface}\label{sec:dft_ASW}

\begin{figure}[ht]
    \centering
    \includegraphics[angle=0, width = 1\linewidth]{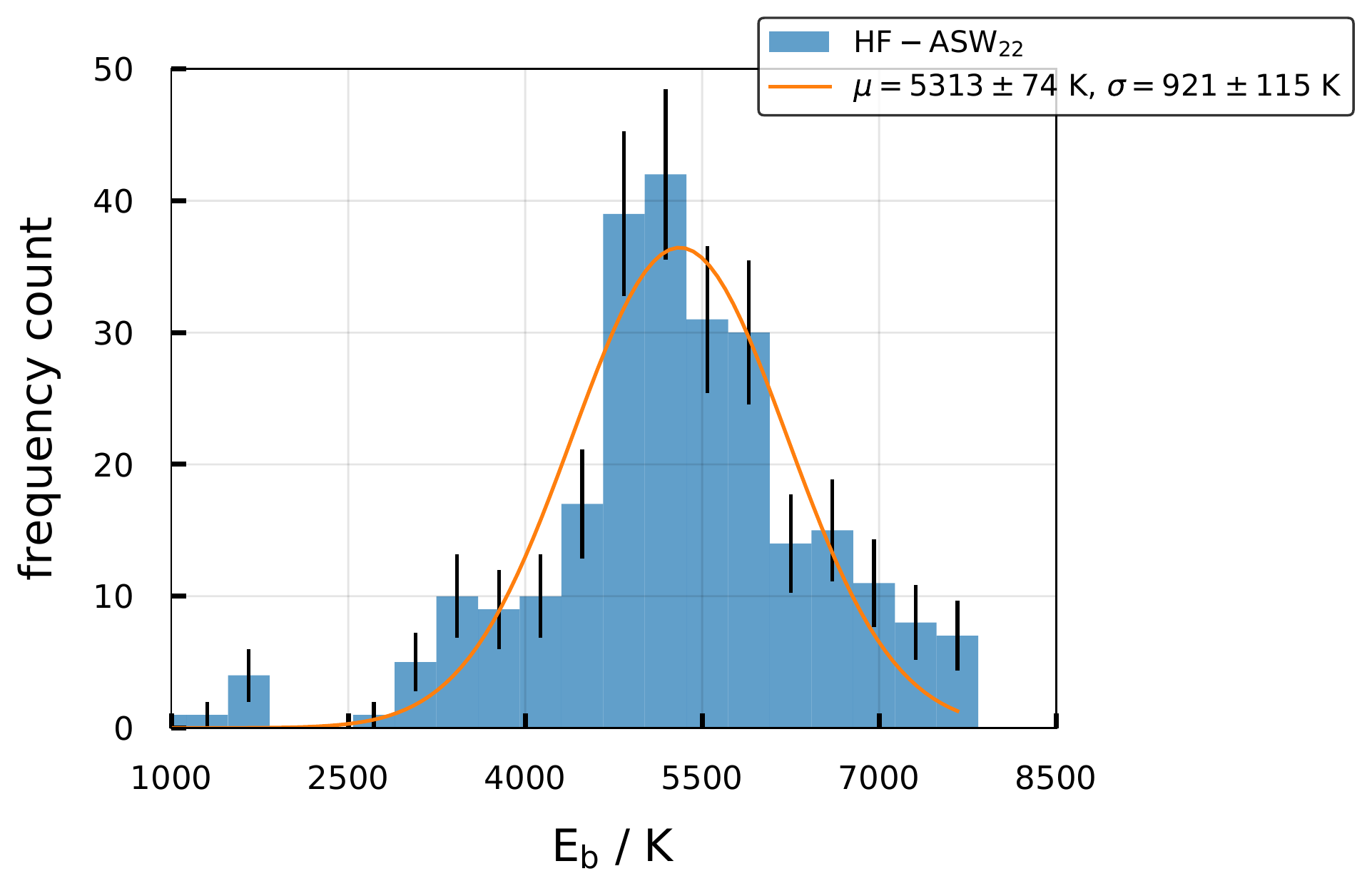}
    \caption{Histogram of the binding energy distribution obtained for the \ce{HF - ASW_22} system.
    To fit the data we employed a bootstrap method (see Sec S3 of the Supplementary Material).
    Mean ($\mu$) and standard deviation ($\sigma$) of the Gaussian fit are reported in the legend.} 

    \label{fig:BE_ASW}
\end{figure}

\begin{figure}[ht]
    \centering
    \includegraphics[angle=0, width = 0.8\linewidth]{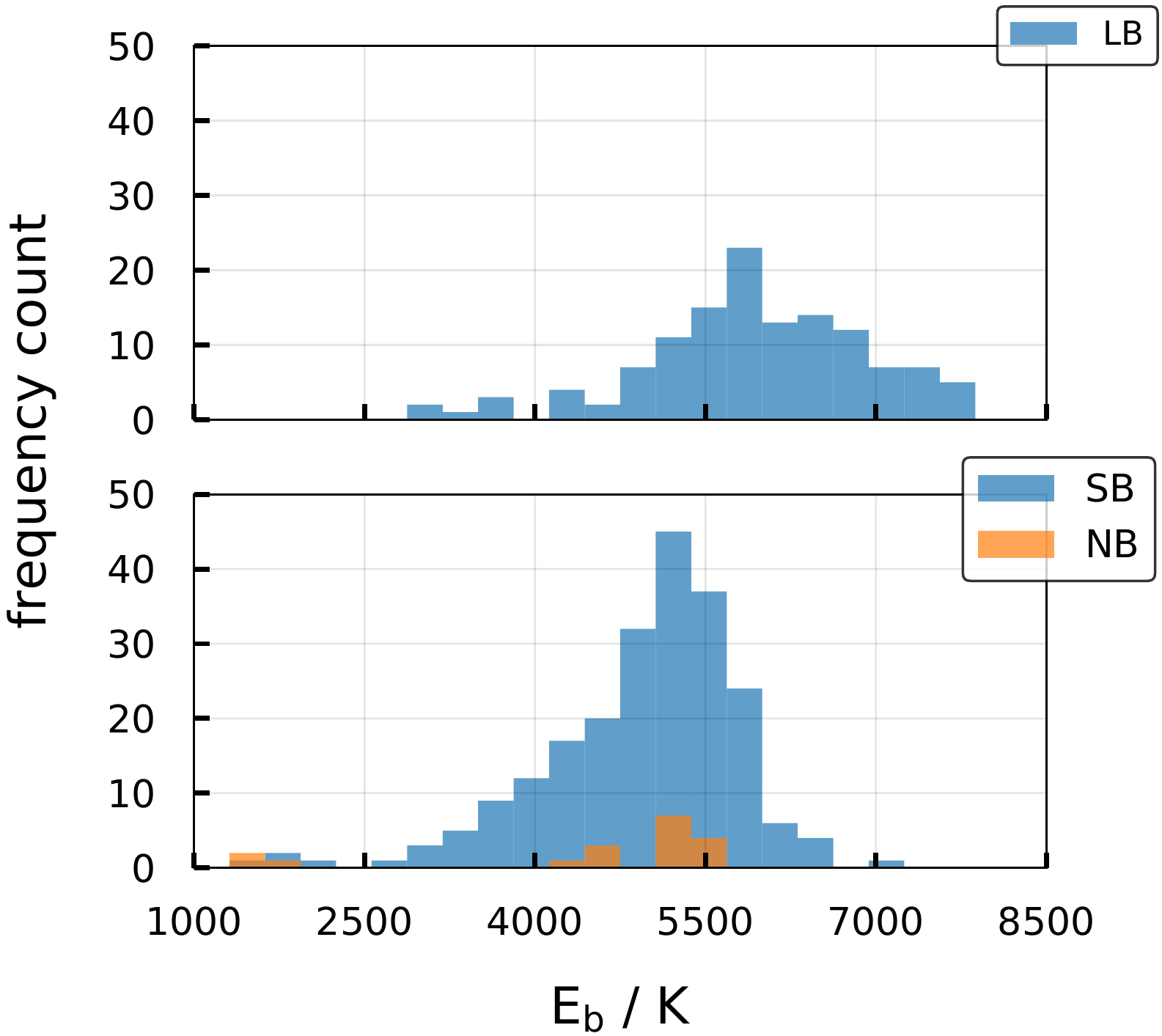}
    \caption{The binding energy distributions obtained for the different adsorption patterns: \textit{long bridge} (LB), \textit{short bridge} (SB), \textit{non-bridged} (NB). (see section \ref{sec:section41} for classification).} 
    \label{fig:BE_modes}
\end{figure}

\noindent The procedure described in Section \ref{sec:surfmodel} and \ref{sec:bindingsites}
provided a total of 255 unique minimum energy structures on 10 selected ASW clusters containing 
22 water molecules (ASW$_{22}$). These represent a wide range of structurally 
different binding sites and produce a binding energy distribution which ranges from 
1400 K to 7841 K, with an average of 5263 K, reported in Fig. \ref{fig:BE_ASW}.
Using the binding modes classification found for the small and intermediate-size clusters 
(see next section) 
we have been able to decompose the distribution in different contributions
(Fig. \ref{fig:BE_modes}).  In Fig. \ref{fig:ASW_fig}, for each binding mode, we show the most energetically 
favorable structures. In Table \ref{tbl:BE_ASW} we report the percentage of binding sites found for each binding mode and the average $E_b$, along with 
ZPVE corrected values.

{}

\subsection*{Binding modes classification}\label{sec:section41}

\begin{figure}[ht]
    \centering
    \includegraphics[angle=0, width = 0.9\linewidth]{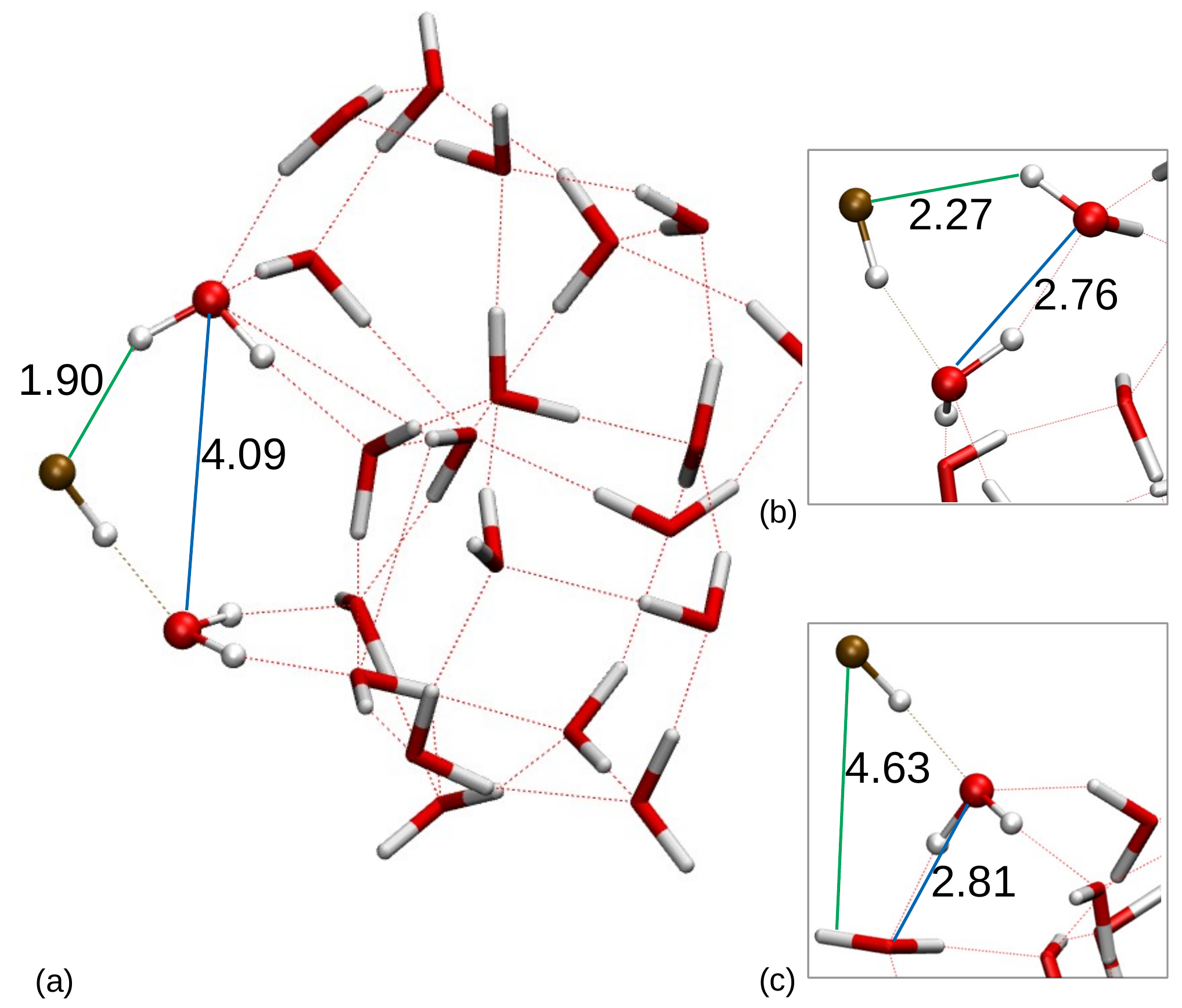}
    \caption{\ce{HF-\ce{ASW_22}} system. Equilibrium geometries calculated at revPBE0/def2-TZVP level. We reported the structures highest in binding energy for each adsorption pattern: (a) \textit{long bridge}, (b) \textit{short bridge}, (c) \textit{non-bridged} ($E_b$ values reported in Table \ref{tbl:BE_ASW}). The atoms within 3\text{\AA} range of F atom are represented as ball and sticks. The figure also shows $R_{F-Hw}$ (green) and $R_{O-O}$ distances (blue). The color scheme for the
atoms is yellow for F, red for O, and white for H.} 
    \label{fig:ASW_fig}
\end{figure}

\begin{figure*}[ht]
    \centering
    \includegraphics[angle=0, width = 0.8\linewidth]{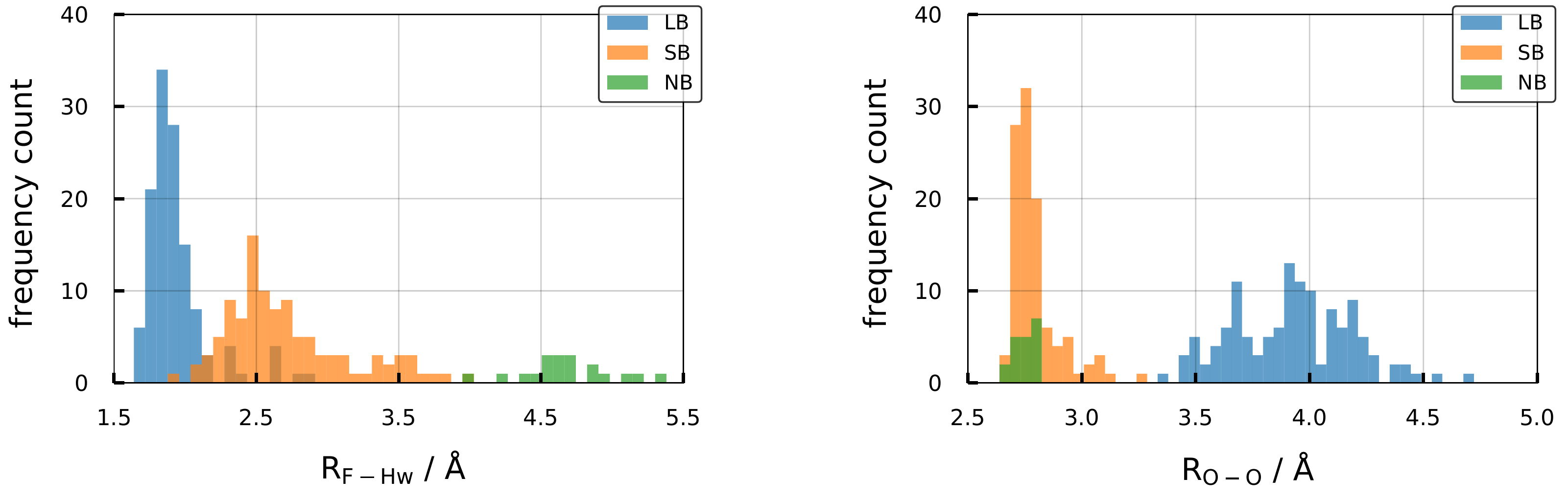}
    \caption{Histograms of the $R_{F-Hw}$ (left panel) and $R_{O-O}$ (right panel) distances detected in the 255 structures considered for HF -- \ce{ASW_22} system. The different adsorption patterns are identified by colors: \textit{long bridge} (blue), \textit{short bridge} (orange), \textit{non-bridged} (green) (see section \ref{sec:section41} for classification).} 
    \label{fig:hist_fig}
\end{figure*}

\noindent Three main different binding adsorption patterns have been found to be recurrent: 
two bridged modes where HF binds to the surface through two different hydrogen bonds, and a non-bridged one where a single hydrogen bond is formed. 
Here we explain the identification procedure we followed, applying it to the
HF -- \ce{ASW_{22}} system. For the first group of structures (bridged), it is possible 
to discriminate between the two different modes through the length of the distance between 
the two oxygen atoms closest to the  binding site ($R_{O-O}$, marked in blue in Fig.
\ref{fig:ASW_fig}). First, a long bridge (abv. LB, Fig. \ref{fig:ASW_fig}a), in which $R_{O-O}$ 
is longer than 3.3 \text{\AA}, and second a short bridge  (abv. SB, Fig. \ref{fig:ASW_fig}b) 
with a $R_{O-O}$ shorter than that distance.  This criterion has been established with the 
aid of the histogram obtained for the aforementioned distance reported in 
Fig. \ref{fig:hist_fig}. The right panel  shows the presence of two distributions, one 
centered around 3.8 \text{\AA} which corresponds to the LB-type binding mode, while a 
second distribution appears at 2.7 \text{\AA}, which can be associated with the SB- and the 
NB-type (the non-bridged, Fig. \ref{fig:ASW_fig}c). Furthermore, the NB-type structures can 
be separated from the SB- ones using $R_{F-Hw}$: the distance of the fluorine-water hydrogen 
bond (Fig. \ref{fig:ASW_fig}, marked in green).  As shown in the left panel of
Fig. \ref{fig:hist_fig}, two distinct distributions appear on the short distance region, 
belonging to the bridged types (LB and SB), thus confirming that they correspond to 
different binding modes.  As expected, the NB-type binding mode presents a
distribution centered at much larger $R_{F-Hw}$ distances, namely around 4.6 \text{\AA}, 
consistent  with the absence of a second hydrogen bond.
Finally, there is a last category that is a sub-classification of the bridged type
structures, in which there are 3 oxygen atoms within a 3 \text{\AA} sphere range from 
the fluorine  atom. The latter forms a double hydrogen bond with two different water 
molecules in a Y-shaped  binding mode. An example of this pattern is constituted
by the minimum energy structures of HF -- \ce{W_{5-6}} (S5a-LBY, S6a-LBY) reported 
in Fig. \ref{fig:w5_w6}.

{}

{}

{}

\section{Discussion}\label{sec:section4}

\subsection{Energetic considerations}
\noindent Our analysis indicates that HF is adsorbed on the small water clusters 
and ASW via a non-covalent interaction. The global energy  minima for the 
HF -- \ce{W_{3-6}} systems correspond to a long bridge-type bond ($E_b$ between 6200 K 
and 7753 K). Hence,  the LB adsorption configuration seems to be energetically very 
favorable. This occurs despite a significant stretching of the water oxygen framework
to attain the bound configuration, which is reflected in high deformation
energies (2976 K on average, see Table \ref{tbl:small clusters} and \ref{tbl:W5_W6}).
However, the concurrent large interaction energies ($\sim$ 9000 K)
exceeds the deformation energy, making it an overall advantageous configuration. On the 
other hand, the deformation of the oxygen framework in the other binding modes is less pronounced, 
albeit at a cost of smaller interaction energies, resulting in lower 
binding energies.  This trend is confirmed in the ASW system. The binding energy 
distributions found for LB and SB modes (Fig. \ref{fig:BE_modes}) are centered 
at 5888 K and 4659 K respectively,  thus confirming the stronger energetic interaction in 
the LB-type mode.  It is worth mentioning that, although in the NB binding mode HF only forms one hydrogen bond with the ASW surface,  the binding energy is, on average, only 300 K smaller than the short bridge-type. However, this binding mode   
is also less frequent representing the 8\% of the equilibrium structures found. 
Finally, the Y-type variant seems to be energetically favorable (largest $E_{int}$ $\sim$ 9820 K), 
which can be attributed to the formation  of a double hydrogen bond of fluorine with two water molecules. 

\subsection{Effect of the ASW environment on the binding energies}\label{sec:section42}

\begin{figure*}[ht]
    \centering
    \includegraphics[angle=0, width = 0.9\linewidth]{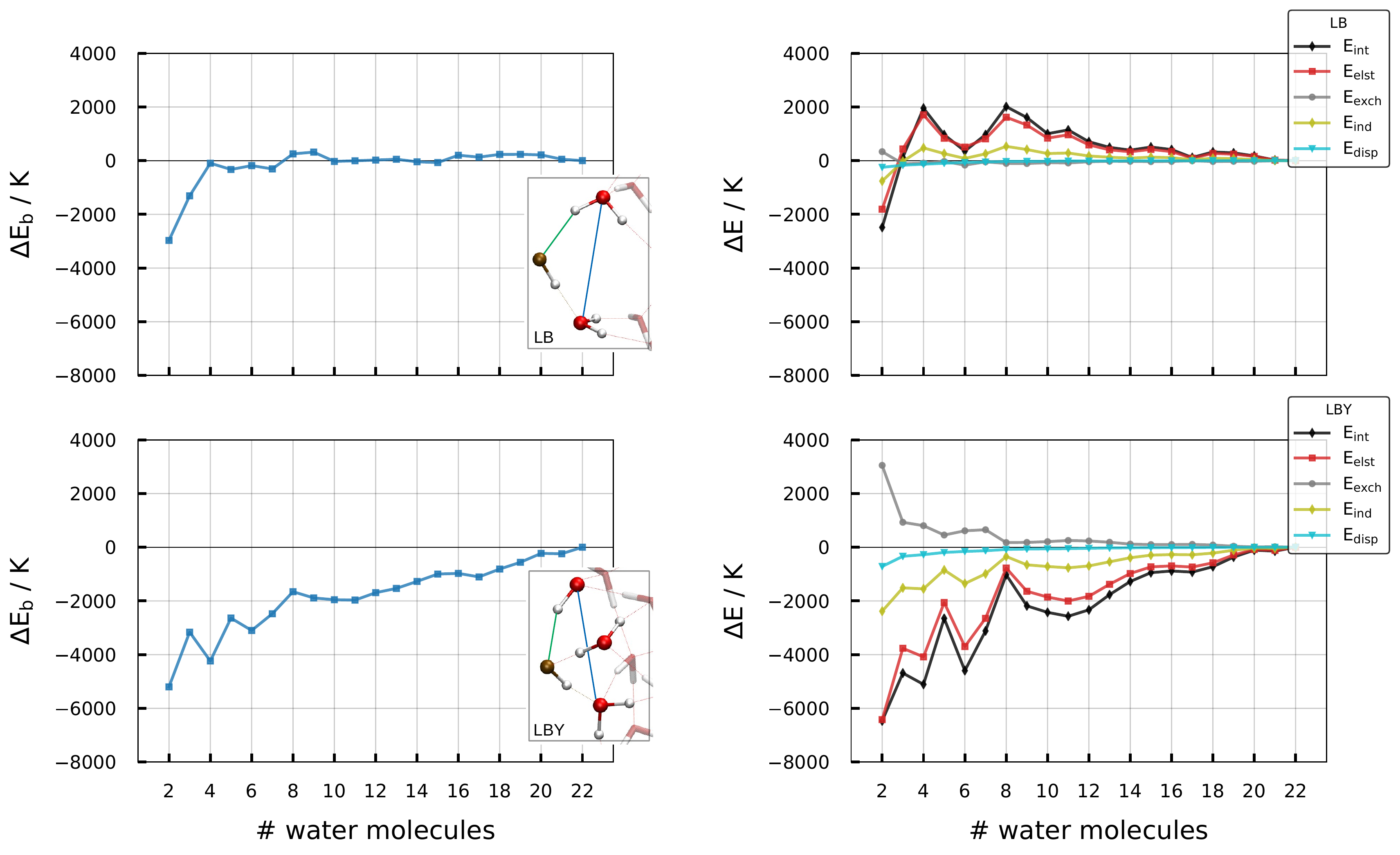}
    \caption{Changes in the binding energy ($\Delta E_b$) (left panel) and the interaction energy 
    ($\Delta E_{int}$) (right panel) with respect to the numbers of water molecules 
    in the frozen \ce{ASW_{22}} cluster in the supermolecular configuration. The interaction energy is further decomposed into electrostatic, induction, dispersion and exchange contribution. 
    The interaction energy is decomposed using SAPT0 analysis. The 0 K baselines correspond to the values in HF --\ce{ASW_{22}}.
    } 
    \label{fig:sapt plot}
\end{figure*}{}

\noindent The results presented in the previous sections suggest that the surface properties 
of the ASW in the proximity of HF binding sites have a significant effect on the binding energies. 
However, there is no correlation between the binding energies and the
$R_{F-Hw}$ or the $R_{H-Ow}$  distances (Fig. S18), thus pointing to a significant energetic influence  of
the electrostatic field produced by the water molecules.
In order to investigate this point, we chose two structures (a LB and a LBY type, specifically the 
highest in binding energy for both groups)  and followed the evolution of the binding energy with respect to 
the size of the cluster, by removing one water molecule at a time, starting from the optimized structure, and keeping the remaining molecules frozen. We repeated the process until all but the two nearest water molecules to the HF binding site are removed. 
The resulting $\Delta E_b$ values, computed for each frozen water cluster, are plotted in Fig. \ref{fig:sapt plot}, left panel.   
The 0 K baseline corresponds to the original energy value for the HF -- \ce{ASW_22} structures, 
such that the plotted energies represent the deviation from that binding energy at each cluster size.
Although the two structures present a  similar $E_b$ (Table \ref{tbl:sapt}), they show a different convergence 
to the  HF -- \ce{ASW_22} final value. When the water clusters consists only of two molecules (\ce{W_2} cluster) 
both the LB and LBY structures  significantly  underestimate the final values. 
However, $\Delta E_b$ of the LB structure computed at \ce{W_4} cluster is already converged, 
which indicates that, for this mode,  only the local water environment has a strong effect 
on the binding energy.   On the other hand, the LBY structure shows a slower
convergence, and reaches values within 500 K of the ASW$_{22}$ binding energy only when the water cluster
grows to 18 molecules. Thus, in this case, the water environment seems to have an important impact on 
the binding energy, as it accounts for approximately 60\% of the target $E_b$ value.

\begin{table}[h!]
     \centering
     \small
     \caption{The different contributions to the interaction energy ($E_{int}$) calculated using SAPT0 for a LB- and a LBY-type equilibrium structure of HF-- \ce{ASW_{22}}. The total binding energies ($E_b$) are also shown. All the energy values are in Kelvin. List of abbreviation used: $E_{elest}$ -- electrostatic energy; $E_{exch}$ -- exchange-repulsive energy; $E_{ind}$ -- induction energy; $E_{disp}$ -- dispersion energy; LB -- \textit{long bridge}, LBY -- \textit{long bridge Y-type}} 
\label{tbl:sapt}
  \begin{tabular*}{0.48\textwidth}{@{\extracolsep{\fill}}lllllll}
    \hline
  &   \multicolumn{1}{c}{revPBE0/} & \multicolumn{5}{c}{SAPT0/} \\
  & def2-TZVP & \multicolumn{5}{c}{jun-cc-pVDZ} \\\cmidrule{2-2} \cmidrule{3-7}   
 Type &  \multirow{1}{*}{$E_b$} & $E_{int}$ &  $E_{elest}$ & -$E_{exch}$ & $E_{ind}$ & $E_{disp}$\\
\hline
    LB & 7841  & 10975 & 15433  & 14962 & 6988 & 2129 \\
    LBY & 7418 & 12950 & 20221 & 19954 & 9858 & 2823 \\
     \hline
\end{tabular*}
\end{table}

To further analyze the environment effect on the interaction energy, we decomposed it 
into electrostatic ($E_{elest}$), exchange ($E_{exch}$), induction ($E_{ind}$) and dispersion ($E_{disp}$) energy, using SAPT0. 
The values for each contribution to the interaction energy are shown in 
Table \ref{tbl:sapt}. As expected,  the largest attractive contribution to the interaction energy 
comes from the electrostatic, followed by induction and dispersion  energy. The only
repulsive interaction corresponds to the exchange energy. All contributions to the 
interaction energy are larger in the LBY structure, which is consistent with a HF molecule
further embedded into the oxygen framework compared to the LB structure. 
Fig. \ref{fig:sapt plot}, right panel, shows the evolution of the different contributions with respect 
to the number of molecules in the  ASW cluster. In the LB-type binding mode, the $E_{exch}$, $E_{ind}$, and 
$E_{disp}$ contributions converge rapidly, not showing any major variations after the third water molecule. 
The $E_{elest}$ contributions fluctuates within a range of -2000 and 2000 K, and reaches a value within 500 K of 
the final value at \ce{W_10} cluster size. Therefore, only the nearest 10 water molecules affect the 
interaction energy and mostly the electrostatic term. The other contributions 
are almost exclusively determined by the two closest water molecules bound to HF.
On the other hand, in the LBY binding mode, both $E_{exch}$ and $E_{ind}$ show
changes up to \ce{W_6} cluster size, especially when adding the third water molecule. This is in line with the 
observation  that, in this mode, HF interacts strongly with the three water molecules closest to the binding site. 
For the $E_{elest}$ the variance with respect to the water environment is even more dramatic, since at \ce{W_2}
it is 6000 K lower than the final value and the convergence is very slow: it gets within 500 K of the \ce{ASW_22} value only at
\ce{W_18}. Thus, in this LBY sample, even distant water molecules play a paramount role in an accurate determination 
of the binding energy value.

\subsection{Effect of the surface topology on the binding energies}
\noindent In order to assess whether the binding energy values obtained for the \ce{ASW_{22}} system are dependent on the cluster size, 
we applied our simulation procedure (Sec. \ref{sec:surfmodel} and \ref{sec:bindingsites})  to two clusters of 37 water molecules. 
The resulting structures are  amorphous and oval-like shaped ($\sim$ $11\times 8 \times 5) \text{\AA}^3 $, 
see Fig. S19 in the Supplementary material. We performed  both the initial minima search and the 
binding site optimizations  at BLYP/def2-SVP level, while the $E_b$ energies have been
computed at revPBE0/def2-TZVP. The procedure provided a total of 126 unique binding sites. In 
Table \ref{tbl:w37} we report the average binding energies, as well as the percentage of each binding mode.
As for the \ce{ASW_{22}} cluster, the most frequent structure type is the LB ($\sim40\%$), 
followed by the SB ($\sim25\%$) while the NB-type is completely absent. Noticeably, the LBY variant 
is particularly common in this system ($\sim30\%$), which constitutes the main difference 
with respect to  the \ce{ASW_{22}} cluster.  Since the geometry for HF -- \ce{ASW_{37}} were computed using a 
GGA level functional and a double zeta basis set, in order to estimate the error  in energy, and being able to compare   
to the \ce{ASW_{22}} system results, we refined the optimization of 15 structures at the higher revPBE0/def2-TZVP level
of theory, and computed new binding energies. We used these values to construct a linear fit to correct all the energies. The resulting fitted binding energy values are displayed in Table \ref{tbl:w37}. The fitted binding energies of the HF -- \ce{ASW_37}  system are in 
good agreement with the values obtained for HF -- \ce{ASW_22}. The average value is only $\sim$ 100 K higher  
in energy and the highest binding energy value is about 330 K above the ones found for the 22 water 
molecules cluster. Therefore, it seems that increasing the  cluster size has only a minor effect on the computed 
binding energy distribution. In conclusion, the \ce{ASW_{22}} model is a good compromise between maximizing the 
number of different binding sites, while still employing a high level of theory.

\begin{figure*}[ht]
    \centering
    \includegraphics[angle=0, width = 0.8\linewidth]{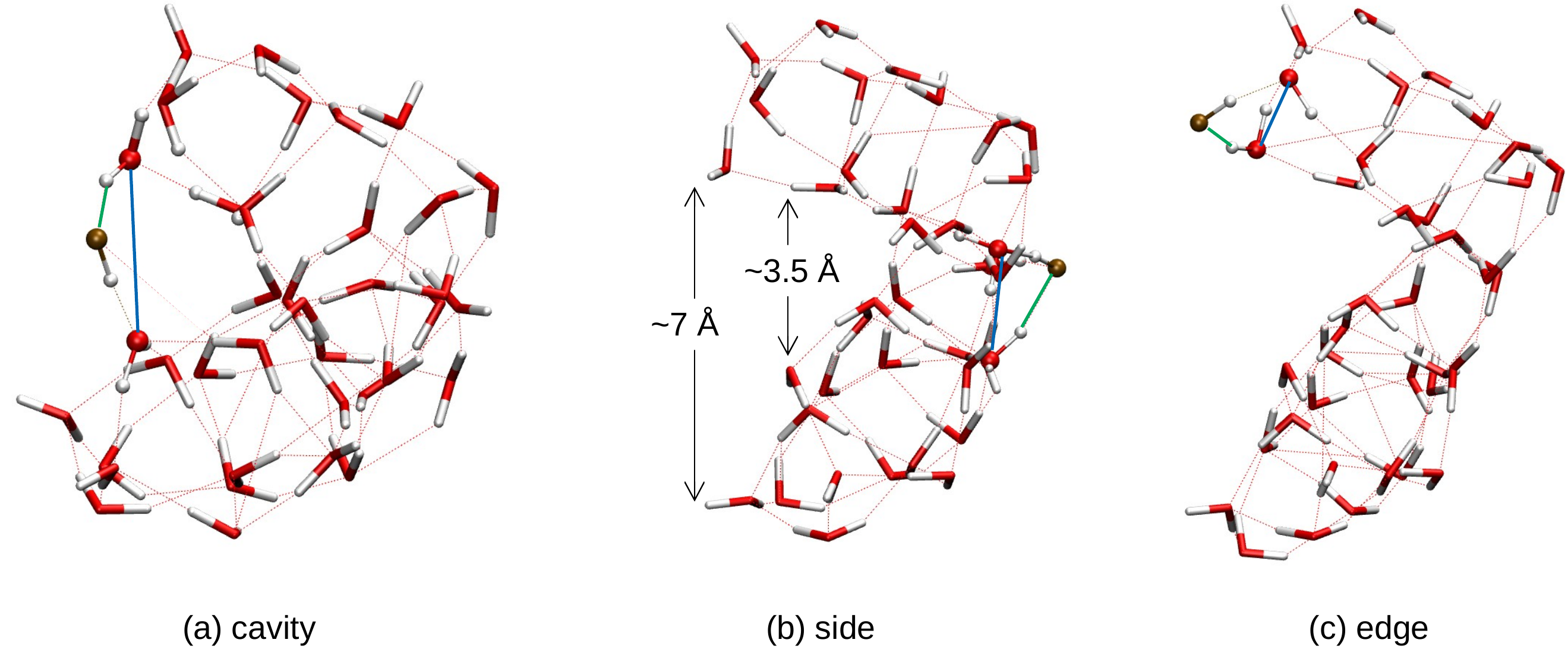}
    \caption{Equilibrium structures of HF--\ce{W37} system, provided of cavity, computed at BLYP-def2-SVP level. We reported three structures of interest: (a) HF inside the cavity; (b) HF on the side; (c) HF on the cavity's edge ($E_b$ values reported in Table \ref{tbl:w37_cavity}). The atoms within 3\text{\AA} range of F atom are represented as ball and sticks. The figure also shows $R_{F-Hw}$ (green) and $R_{O-O}$ distances (blue). The color scheme for the
atoms is yellow for F, red for O, and white for H.} 
    \label{fig:cavity}
\end{figure*}{}

Finally, motivated by recent works, pointing to an increase of the binding energy induced by the presence of cavities on 
the surface \citep[see][]{rimola2018,enrique-romero2019}, we decided to further investigate the influence of the shape of the ASW surface considered.
We constructed an \textit{ad hoc} \ce{ASW_{37}} system (see Fig. \ref{fig:cavity})
incorporating a wedge-like cavity, by putting together two of the 22 molecule clusters\footnote{Note that seven molecules have been removed to made the junction modelling easier; the cavity is of around 10 \text{\AA} length, 3.5 \text{\AA} average width and a maximum width of 7 \text{\AA} 
reached at the edges level.}. 
After the surface sampling procedure, we computed the binding energy for some representative structures.

\begin{table}[h!]
     \centering
     \small
     \caption{The average binding energy including BSSE counterpoise correction for the system \ce{HF - ASW_37}. All the energy values are in Kelvin. List of abbreviation used: SB -- \textit{short bridge}, LB -- \textit{long bridge}, NB -- \textit{non-bridged}, Y -- Y type. Energy values in kcal $\mathrm{mol^{-1}}$ are reported in Supplementary Material}
     \label{tbl:w37}
  \begin{tabular*}{0.48\textwidth}{@{\extracolsep{\fill}}lllll}

    \hline
 HF -- \ce{ASW_37} & & \multicolumn{3}{c}{revPBE0/def2-TZVP} \\ \cmidrule{3-5}  
 
 & & Highest & \multicolumn{2}{c}{Average} \\ \cmidrule{3-3} \cmidrule{4-5}
 Type & \%  & \multirow{1}{*}{$E_b$} & \multirow{1}{*}{$E_b$} & $\sigma$  \\
\hline
     Tot & 100 & 8170 & 5369 &   1283 \\
    LBY & 30.16 & 8170 & 5326 &  1546 \\
    LB & 39.68 & 7903 & 5772 &   1128 \\
    SBY & 5.55 & 7641 & 5548 &   1488 \\
    SB & 24.60 & 5968 & 4680 &   720 \\
    \hline
\end{tabular*}
\end{table}

\begin{table}[h!]
\small
     \centering
     \caption{The binding energies including BSSE counterpoise correction for the system \ce{HF - ASW_37} containing a  cavity. 
     All the values are in Kelvin. List of abbreviation used: $E_{def}$ -- deformation energy; LB -- \textit{long bridge}, Y -- Y type. Energy values in kcal $\mathrm{mol^{-1}}$ are reported in Supplementary Material}
     \label{tbl:w37_cavity}
  \begin{tabular*}{0.48\textwidth}{@{\extracolsep{\fill}}l cc}

    \hline
 HF -- \ce{ASW_37} / cavity &   \multicolumn{2}{c}{revPBE0/def2-TZVP} \\ \cmidrule{2-3}    
 & $E_b$ & $E_{def}$  \\
\hline
    cavity LB & 9196  & 490  \\
    side LBY & 7044  & 3201  \\
    edge  LB & 6943  & 968  \\
    \hline
\end{tabular*}
\end{table}

We explored three different spatial situations for the HF molecule: 
inside the cavity, outside of it and on the edge (panels a, b, and c in Fig. \ref{fig:cavity}, respectively). In Table \ref{tbl:w37_cavity} we report 
the computed values, applying the fitting correction previously described in order to account for the geometry error.
In line with the results by \citet{enrique-romero2019}, we find larger $E_b$ values in 
the cavity, compared to the sites on the side and on the cavity's edge  (energy difference around 2000 K); even though 
the number of samples considered is not sufficient to establish a proper statistical analysis.
The reason for the higher values seems to be a low cluster deformation energy (see Table \ref{tbl:w37_cavity}),
such that  energetically favourable binding sites
are more readily available to the HF molecule, compared to an ASW surface without cavities. 
It is therefore likely that a rugged ASW surface 
might shift the distribution to higher energies, but 
a more detailed investigation is needed to confirm this trend.

\subsection{Comparison with previous works}

\noindent Despite its diagnostic potential as interstellar gas tracer, HF molecule had not been 
theoretically studied in a extensive way heretofore. Binding energy values with respect to the
interaction of HF with different type of grain surface can be found in literature: hydrogenated crystalline silica grain surface  \citep{vanderwiel2016}, and amorphous silica
surface covered by different types of ices, specifically \ce{H2O} \citep{vanderwiel2016},
\ce{CO} \citep{rivera-rivera2012a}, \ce{CO2}  \citep{chen2006}). To the best of our
knowledge, \citet{das2018} binding energy relatively to the system HF -- \ce{W_4}
is the only one provided for a water cluster. They computed a value of 5540 K at MP2/aug-cc-pVDZ
(counterpoise correction and ZPVE correction not included), that falls within the range of energies
we found for the same  system (6211-1208 K). Specifically, their structure corresponds to our S4d-SB minimum (5338 K) 
that is not the global minimum on that potential energy surface.
It is also worth pointing out, that the value has been obtained without performing a previous minima search, therefore, apart from the low level of model chemistry in \citet{das2018}, 
the existence of multiple binding sites is not considered. 
The other value for HF on water surfaces that has hitherto been used in astrochemical models,
is the one obtained from the semi-empirical approach by \citet{Wakelam2017}.
The model consists of a linear fit between the binding energies on water monomers and experimental
values on ASW surfaces.  The binding energy estimated using this model is 7500 K, which significantly 
overestimates the average value of 5490 K at the peak of our Gaussian distribution. Nonetheless it falls 
within the binding energy range, being slightly lower than the global minimum (7841 K) on  ASW$_{22}$. 

\subsection{Impact on astrochemical models}
\noindent The desorption process in astrochemical models is 
usually treated by following a classic Polanyi-Wigner approach,  with an exponential dependence on 
the binding energy. This means that a difference of a few tens of Kelvins in the binding energy can 
dramatically affect the fate of the molecule in gas-phase and its final abundance.
More importantly, in line with the results of this work, experimental studies \citep{amiaud2006,He2011,Noble2012} 
have found a distributions exploring the relationship between desorption energy and degree of 
surface coverage 
(i.e the number of  molecule presents on a surface respect to the monolayer deposition). Even the 
analysis carried out at low coverage, resembling ISM conditions of very low density, leads to the 
presence of a binding energy distribution, thus, it seems to be required to reconsider the use of a single binding energy 
value, employed in current astrochemical models \citep{grassi2020a}. We expect a high impact on the surface chemistry, relevant 
for studies related to the formation of complex organic molecules in star-forming regions. The 
presence of a binding energy distribution, in fact, will allow a richer set of available sites with a 
variety of desorption energies, which will increase the reactivity between different molecules (via 
thermal hopping or tunneling) due the longer residence time of the molecules on the surface. The 
consequence might have a dramatic effect on the amount of species returning to  gas phase, capable to 
be observed through their rotational lines.

\section{Conclusions}\label{sec:section5}
\noindent In summary, in this paper we presented a new procedure to compute binding energy distributions
on ASW model surface. The pipeline consists of three steps: 1) An extensive DFT benchmark on 
small water clusters, using high-level wave-function methods as a reference, to obtain an adequate
DFT functional. 2) \textit{Ab initio} MD simulation of water to obtain an ASW surface model. 3) Sampling of the binding
sites on the surface with the target molecule and computation of the binding energies. 
Applying the simulation procedure to the hydrogen fluoride molecule, we found 255 unique structures
corresponding to 3 different binding modes. The center of the distribution is located at $5313\pm74$ K ($10.56\pm0.15$ kcal $\mathrm{mol^{-1}}$). The binding mode 
that exhibits the highest energy values, corresponds to HF bound to the ASW surface through two different 
hydrogen bonds. The extended water environment (beyond the two nearest water molecules) seems to
have a significant impact on the binding energy for some of the equilibrium structures, especially on the
electrostatic component of the interaction energy. In the future we plan to use this pipeline to build a binding
energy distribution database of many relatively small molecules on ASW and other astrophysically relevant surfaces.

\section*{Acknowledgements}
\noindent GB gratefully acknowledges support from Sapienza Università di Roma fellowship D.R. 1053/2018 Prot. n. 0030665; Beca de Doctorado UCO 1866 and Beca de Doctorado Nacional ANID n. 21200180.
SB is financially supported by ANID Fondecyt Iniciaci\'on (project 11170268), and BASAL Centro de Astrofisica y Tecnologias Afines (CATA) AFB-17002.
SVG is financially supported by ANID Fondecyt Iniciaci\'on (project 11170949)   .

\section*{Supplementary materials}
\noindent Supplementary material associated with this article, including all optimized xyz geometries, 
can be found, in the online version.

\bibliography{Astrochem,wf-methods,ISM,Basis_sets,DFT_functionals,qc_programs}

\end{document}